\documentclass[10pt, conference, compsocconf]{IEEEtran}
\usepackage{amsmath, amssymb}
\usepackage{algorithm}
\usepackage{algorithmic}
\usepackage[dvips]{graphicx}

\begin{document}
\title{Distributed Greedy Scheduling for Multihop Wireless Networks}

\author{\IEEEauthorblockN{Albert Sunny}
\IEEEauthorblockA{Centre for Electronics Design and Technology \\
Indian Institute of Science \\
Bangalore-560012\\
salbert@cedt.iisc.ernet.in}
\and
\IEEEauthorblockN{Joy Kuri}
\IEEEauthorblockA{Centre for Electronics Design and Technology \\
 Indian Institute of Science \\
  Bangalore-560012 \\
kuri@cedt.iisc.ernet.in}
}

\maketitle

\begin{abstract}
We consider the problem of scheduling in multihop wireless networks subject to interference constraints. We consider a graph based representation of wireless networks, where scheduled links adhere to the $K$-hop link interference model. We develop a distributed greedy heuristic for this scheduling problem. Further, we show that this distributed greedy heuristic computes the exact same schedule as the centralized greedy heuristic.
\end{abstract}


\IEEEpeerreviewmaketitle

\section{\large{Introduction and Related Work}}
\label{intro}
Scheduling and routing algorithms allocate resources to competing flows in multihop wireless networks. Research into scheduling, routing and congestion control is several decades old, but has seen a lot of activity , following the seminal paper of Tassiulas and Ephremides \cite{tassiulas}. One possible way to schedule links in a wireless network is to use a spatial time division multiple access (STDMA) along with the physical interference model. While physical interference model allows more aggressive scheduling, it has been shown that no localized distributed algorithm can solve the problem of building a feasible schedule under this model \cite{brar}. Since the paper by Kumar and Gupta \cite{gupta}, the protocol model of wireless network has been studied extensively. Research has shown that a $K$-hop link interference model can be used to effectively model the protocol model \cite{sharma}. 

A commonly used model is the $K$-hop link interference model, in which two links that are not within K-hops of each other can communicate simultaneously, and the capacity of a link is a constant value if there is no interference \cite{lin}, \cite{chaporkar}, \cite{wu}, \cite{joo}, \cite{lin1}, \cite{wu1}. In \cite{lin}, the Maximal Matching (MM) scheduling algorithm is used under the node exclusive interference model. This algorithm can operate in a distributed fashion and is proven to achieve at least one half of the achievable throughput. This has motivated subsequent research on distributed algorithms with provable performance \cite{chaporkar}, \cite{wu}, \cite{wu1}, \cite{joo}, \cite{lin1}.

Scheduling algorithms under different SINR interference models have been studied in the literature \cite{neely}, \cite{chiang}, \cite{soldati}, \cite{gro}, \cite{kim}. In \cite{neely}, the authors have proposed a simple and distributed scheduling algorithm, that is an approximation to the optimal centralized algorithm. In \cite{chiang}, for the logarithmic SINR interference model, the author has proposed a distributed algorithm that is distributed and optimal when SINR values are high. The authors in \cite{soldati} and \cite{gro} have also proposed heuristic algorithms under the target SINR interference model where the capacity of a link is a constant value when the received SINR exceeds a threshold, or zero otherwise. In \cite{kim}, the authors have explore localized distributed scheduling for linear and logarithmic SINR model.

The problem of link scheduling under the $K$-hop link interference model has been shown to be NP-hard in \cite{sharma}, \cite{jain}. Motivated by this, we explore heuristics to address the link scheduling problem. In particular, it is interesting to explore the greedy heuristic because it lends itself to a distributed implementation \cite{sharma}. While the idea of a distributed version of the greedy heuristic seems trivial, to the best of our knowledge, it has not been described precisely in the literature,. We find that the distributed greedy heuristic involves certain subtleties, that makes the algorithm non-trivial.

The main results of the paper can be summarized as follows: 
\begin{itemize}
\item We develop a distributed greedy heuristic for the scheduling problem under $K$-hop link interference model. 
\item We prove that the distributed greedy and the centralized greedy scheduling heuristics, compute identical schedules.
\end{itemize}

The remainder of this paper is organized as follows. The system model and the problem formulation are described in Section 2. In Section 3, we describe and analyze the distributed greedy algorithm. We conclude in Section 4.

\section{\large{System Model and Problem Formulation}}
\label{problem}
We represent the network as a directed graph $\mathcal{G=(N,L)}$, where $\mathcal{N}$ represents the set of nodes and $\mathcal{L}$ represents the set of links in the wireless network. We assume that all the wireless transmissions use the same wireless channel and hence interfere with one other. We also assume that all the transmissions happen at a fixed power level (which can be different for different nodes).  A wireless link $(i,j) \in \mathcal{L}$ if node $j$ can receive packets from node $i$, provided no other transmissions are going on. The wireless links are considered as directed edges. If we consider a link $(i,j) \in \mathcal{L}$, then we define node $i$ as the source and node $j$ as the sink. If a link has node $i$ as its source node, then that link is called an attached link of node $i$. We note that no matter where the schedule is computed in the network, it needs to be conveyed to the source node of the scheduled link.

IEEE 802.11 based interference model is used for modeling inter-link interference. Here we reproduce some definitions from \cite{sharma} in order to define the interference model. Let $d_S(x,y)$ denote the shortest distance (in terms of number of links) between nodes $x,y \in \mathcal{N}$. Define a function $d\ :\ (\mathcal{L},\mathcal{L}) \to \mathbb{N}$ as follows: For $l_u=(u_1,u_2),\ l_v=(v_1,v_2)\in \mathcal{L}$, let
$$d(l_u,l_v)=\min_{i,j\in{1,2}}d_S(u_i,v_j)$$
In the $K$-hop link interference model, we assume that any two links $l_1$ and $l_2$ for which $d(l_1,l_2)<K$, will interfere with each other and hence can not be active simultaneously. 

A set of links $\mathcal{M}$ is a maximal independent set, provided no two links of $\mathcal{M}$ interfere with each other under the given interference model, and no other link can be added to $\mathcal{M}$ without violating an interference constraint. The scheduling problem can then be stated as:
\begin{eqnarray}
\max \quad && \sum_{l \in \mathcal{M}}{\lambda_l} \\
\textbf{Subject to:} && \mathcal{M} \in \mathcal{I}_K
\end{eqnarray}
where $\lambda_l$ is the price of link $l \in \mathcal{L}$. We note that the price of each link $l$ is a arbitrary positive number which can characterize various factors and $\mathcal{I}_K$ denotes the set of all maximal independent sets possible under the $K$-hop link interference model. We assume all nodes have their clocks synchronized to a global time, within a reasonable degree of accuracy. We also assume that there is a reliable mechanism, to pass message between nodes.

\section{\large{Distributed Greedy Heuristic}}
\label{work}

The main intention behind selecting the greedy heuristic as a scheduling policy is that it can be implemented in a distributed manner. Here, we present the Greedy Heuristic as in \cite{sharma}.

\begin{algorithm}
\caption{Centralized Greedy Heuristic}
\begin{algorithmic}[1]
\STATE $W:= \phi$ and $i := 1$.
\STATE Arrange links of $\mathcal{L}$ in descending order of price, starting with $l_1, l_2,...  .$ 
\STATE If $W \cup l_i$ is a valid $K$-matching, then, $W := W \cup l_i$ , $i = i + 1$.
\STATE Repeat Step 3 for all links in $\mathcal{L}$.
\end{algorithmic}
\end{algorithm}
\noindent
Here, a set of edges $W$ is a $K$-valid matching if $\forall l_1, l_2 \in W$ with $l_1 \neq l_2$, we have $d(l_1, l_2) \geq K$.

Here we describe the algorithm for the greedy heuristic. This algorithm is implemented at every node $n \in \mathcal{N}$, as given below.

\begin{figure}[h]
\centering
\includegraphics[width=3.2in,height=1in]{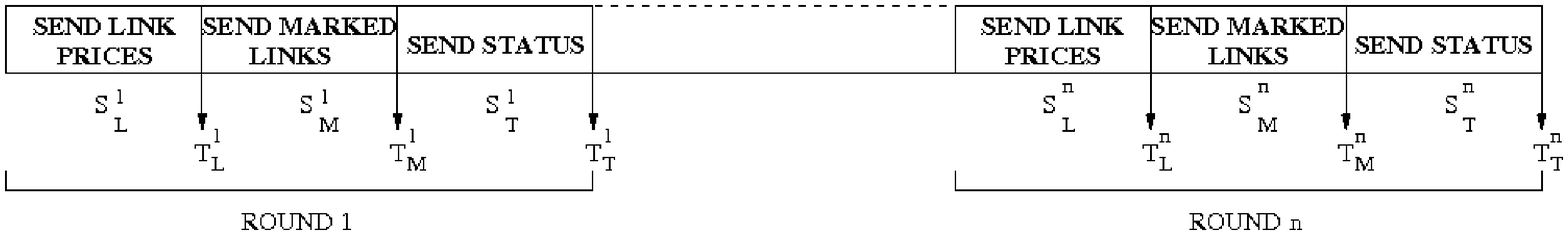}
\caption {Slot division of the distributed greedy algorithm.}
\label{fig:dp}
\end{figure}

The algorithm is laid out as what messages are exchanged between nodes in each slot of the $m^{th}$ \textit{ROUND} and what decision needs to be made at the slot boundary, after the completion of the given slot. The \textit{SEND LINK PRICES} slot, \textit{SEND MARKED LINK} slot and \textit{SEND STATUS} slot, together constitute a \textit{ROUND}. All computations are performed by the nodes themselves, at the slot boundaries, with the local information obtained in the slot immediately preceding the slot boundary.

Let $S^m_L,S^m_M,S^m_T$ be the \textit{SEND LINK PRICES} slot, \textit{SEND MARKED LINK} slot and \textit{SEND STATUS} slot respectively during the $m^{th}$ \textit{ROUND}. Let $T^m_L,T^m_M,T^m_T$ be the slot termination time for the \textit{SEND LINK PRICES} slot, \textit{SEND MARKED LINK} slot and \textit{SEND STATUS} slot respectively, which take part in the the $m^{th}$ \textit{ROUND}. 

Each link in the network can be in any of four states \textit{OPEN (O)}, \textit{CHECK (CH)}, \textit{MARKED (M)} and \textit{CLOSED (CL)}. All links are initially set to \textit{OPEN} and the algorithm set to \textit{DO NOT TERMINATE}.

\begin{algorithm}
\caption{Pseudo-code for Distributed Greedy Heuristic}
\begin{algorithmic}
\STATE \textit{\textbf{In slot $S^m_L$}}
\end{algorithmic}
\begin{algorithmic}[1]
\STATE Disseminate the highest \textit{OPEN} attached link price to $(K+1)$-hop neighbourhood.
\end{algorithmic}
\begin{algorithmic}
\STATE \textit{\textbf{At time $T^m_L$}}
\end{algorithmic}
\begin{algorithmic}[1]
\IF{at least one attached link is \textit{OPEN}}
	\STATE sort the attached \textit{OPEN} links in descending order of link price. Let $l^{'}_{max}$ be the maximum priced link among the attached \textit{OPEN} links.
	\IF{no \textit{OPEN} link prices are received}
		\STATE link $l^{'}_{max}$ is \textit{MARKED} and all other \textit{OPEN} attached links are \textit{CLOSED}, go to 17.
	\ELSE
		\STATE sort received \textit{OPEN} link prices in descending order of link price. Let $l_{max}$ be the maximum priced link among the received \textit{OPEN} links. 
	\ENDIF
	\IF{$(\lambda_{l^{'}_{max}} > \lambda_{l_{max}} )$}
		\STATE link $l^{'}_{max}$ is \textit{MARKED} and all other \textit{OPEN} attached links are \textit{CLOSED}.
	\ELSE
		\FORALL{\textit{OPEN} attached link $l$}
			\IF{$(d(l, l_{max}) < K) $} 
			\STATE link $l$ is set to \textit{CHECK}.
			\ENDIF
		\ENDFOR
	\ENDIF
\ENDIF	
\end{algorithmic}
\end{algorithm}

The links that are \textit{MARKED}, are the maximum-priced links in their corresponding $(K+1)$ hop neighbourhoods. Also the links that are moved into \textit{CLOSED} state will definitely have a \textit{MARKED} link within $K$ hop link distance. These links will continue to remain in their respective states, and will not participate in price dissemination in the subsequent \textit{ROUND}s.

\begin{algorithm}
\begin{algorithmic}
\STATE \textit{\textbf{In slot $S^m_M$}}
\end{algorithmic}
\begin{algorithmic}[1]
\IF{any of the attached links is \textit{MARKED}}
	\STATE disseminate this information to $(K+1)$-hop neighbourhood. 
\ENDIF
\end{algorithmic}
\begin{algorithmic}
\STATE \textit{\textbf{At time $T^m_M$}}
\end{algorithmic}
\begin{algorithmic}[1]
\FOR{each \textit{CHECK} attached link $l$}
	\IF{$(d(l,\textrm{received \textit{MARKED} link})< K))$ for at least one received \textit{MARKED} link}
		\STATE link $l$ is \textit{CLOSED}.
	\ELSE
		\STATE link $l$ remains in \textit{CHECK} state. 
	\ENDIF
\ENDFOR
\STATE	\textit{OPEN} the highest priced attached \textit{CHECK} link.
\STATE Algorithm status is set to \textit{TERMINATE} at nodes which have no \textit{OPEN} or \textit{CHECK} links.
\end{algorithmic}
\end{algorithm}

A link is moved to \textit{CHECK} state, if it sees a higher priced interfering link during price dissemination, but is unable to decide if that link will get \textit{MARKED}. In this slot, \textit{CHECK} links get to know if there is indeed a higher priced \textit{MARKED} links interfering with it. If so, they are \textit{CLOSED}.

\begin{algorithm}
\begin{algorithmic}
\STATE \textit{\textbf{In slot $S^m_T$}}
\end{algorithmic}
\begin{algorithmic}[1]
\IF{at least one attached link is \textit{OPEN} or \textit{CHECK}}
	\STATE send a \textit{DO NOT TERMINATE} message to all nodes in a $(K+1)$-hop neighbourhood.
\ELSIF{got a \textit{DO NOT TERMINATE} message}
    \STATE send a \textit{DO NOT TERMINATE} message to all nodes in a $(K+1)$-hop neighbourhood.
\ENDIF
\end{algorithmic}
\begin{algorithmic}
\STATE \textit{\textbf{At time $T^m_T$}}
\end{algorithmic}
\begin{algorithmic}[1]
\IF{no \textit{DO NOT TERMINATE} message is received}
	\STATE the algorithm has terminated, schedule all \textit{MARKED} links.
\ELSE
	\STATE go to the $(m+1)^{th}$ \textit{ROUND}.
\ENDIF
\end{algorithmic}
\end{algorithm}

The local termination condition is that no attached link is in \textit{OPEN} or \textit{CHECK} state. In the above slot, this information is conveyed to all the other nodes in the network in a distributed manner. This makes sure that the algorithm terminates in a synchronous fashion at each node.
 
Let us see some examples which illustrate the scheduling algorithm. Let us consider a linear network with 7 nodes, with a 2-hop link interference. The link states is shown against the time when different decision are made. For data transfer, only links (1,2), (2,3), (3,4), (4,5), (5,6) and (6,7) are considered; but control traffic can flow in the opposite direction too.

\begin{figure}[h]
\centering
\includegraphics[width=2.5in]{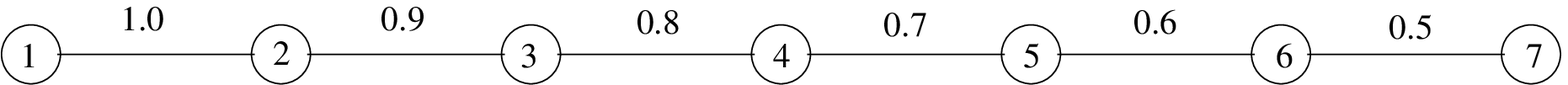}
\caption{Example 1}
\label{fig:example1}
\end{figure}
\begin{center}
\begin{tabular}{|c|c|c|c|c|c|c|}
\hline T & (1,2)  & (2,3) & (3,4) & (4,5) & (5,6) & (6,7)  \\ 
\hline $0$ & \textit{O} & \textit{O} & \textit{O} & \textit{O} & \textit{O} & \textit{O} \\ 
\hline $T^1_L$ & \textit{M} & \textit{CH} & \textit{CH} & \textit{CH} & \textit{CH} & \textit{CH} \\ 
\hline $T^1_M$ & \textit{M} & \textit{CL} & \textit{CL} & \textit{O} & \textit{O} & \textit{O} \\ 
\hline $T^2_L$ & \textit{M} & \textit{CL} & \textit{CL} & \textit{M} & \textit{CH} & \textit{CH} \\ 
\hline $T^2_M$ & \textit{M} & \textit{CL} & \textit{CL} & \textit{M} & \textit{CL} & \textit{CL} \\ 
\hline 
\end{tabular}
\end{center} 

In the first \textit{ROUND}, only link (1,2) is \textit{MARKED}. All other links see a higher priced interfering link and thus move into \textit{CHECK} state. Link (1,2) announces it is \textit{MARKED}. Links (2,3) and (3,4) are \textit{CLOSED}, on reception of this  information, since they interfere with link (1,2). But all other links are moved to \textit{OPEN}, since they do not find any interfering \textit{MARKED} link. This process now repeats itself until the network has no \textit{OPEN} or \textit{CHECK} links.

\begin{figure}[h]
\centering
\includegraphics[width=2.5in]{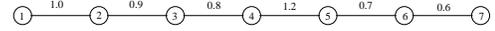}
\caption{Example 2}
\label{fig:example2}
\end{figure}
\begin{center}
\begin{tabular}{|c|c|c|c|c|c|c|}
\hline T & (1,2)  & (2,3) & (3,4) & (4,5) & (5,6) & (6,7)  \\
\hline $0$ & \textit{O} & \textit{O} & \textit{O} & \textit{O} & \textit{O} & \textit{O} \\ 
\hline $T^1_L$ & \textit{O} & \textit{CH} & \textit{CH} & \textit{M} & \textit{CH} & \textit{CH} \\ 
\hline $T^1_M$ & \textit{O} & \textit{CL} & \textit{CL} & \textit{M} & \textit{CL} & \textit{CL} \\ 
\hline $T^2_L$ & \textit{M} & \textit{CL} & \textit{CL} & \textit{M} & \textit{CL} & \textit{CL} \\ 
\hline $T^2_M$ & \textit{M} & \textit{CL} & \textit{CL} & \textit{M} & \textit{CL} & \textit{CL} \\ 
\hline 
\end{tabular}
\end{center} 

In this example, the highest price link is located in the middle of the network. As a result, more links are moved into the \textit{CLOSED} state after the first \textit{ROUND}. And the only remaining \textit{OPEN} link i.e link (1,2), is \textit{MARKED} in the subsequent \textit{ROUND}. 

One can easily compute the centralized greedy schedule and verify that the distributed greedy schedule matches matches it. And we can see the advantage of the \textit{CHECK} state as it prevents a link from getting \textit{CLOSED} after it sees a higher priced link within $K$-hop but which itself may not get \textit{MARKED}. Hence such links move into the \textit{OPEN} state, if no higher priced interfering link has been \textit{MARKED}; else it will move into \textit{CLOSED} state. 

Next, we show analytically that the distributed greedy heuristic schedules the same set of links as the centralized one. \\

\noindent
Let $\mathcal{L}^m_O$  be the set of \textit{OPEN} links before \textit{ROUND} $m$. \\
\noindent
Let $\mathcal{L}^m_C$  be the set of \textit{CLOSED} links before \textit{ROUND} $m$. \\
\noindent
Let $\mathcal{L}^m_H$  be the set of \textit{CHECK} links before \textit{ROUND} $m$. \\
\noindent
Let $\mathcal{L}^m_M$  be the set of \textit{MARKED} links before \textit{ROUND} $m$. \\

\newpage

\newtheorem{lem}{\noindent Lemma}
\begin{lem}
The algorithm terminates in finite time.
\end{lem}

\begin{IEEEproof}
Let $$l^m = arg \max_{l \in \mathcal{L}^m_O} \lambda_l$$ be the global maximum-priced link before \textit{ROUND} $m$. Since the price of this link  among all the \textit{OPEN} attached links is the highest, it is also the local maximum among the \textit{OPEN} attached links received from the $(K+1)$-hop neighbourhood. Thus link $l^m$ gets \textit{MARKED}. Since a \textit{MARKED} link will always remain in the same state, 
$$if\ l \in \mathcal{L}^m_M,\ then\ l \in \mathcal{L}^k_M, \forall k \geq m+1$$
$$ \Longrightarrow \mathcal{L}^m_M  \subseteq \mathcal{L}^{m+1}_M $$

\noindent
Now let us consider link $l^m$, 
$$ l^m \in \mathcal{L}^m_O \Longrightarrow l^m \notin \mathcal{L}^m_M$$

\noindent
But from the previous argument, link $l^m$ gets \textit{MARKED} in \textit{ROUND} $m+1$. Thus
$$ l^m \in \mathcal{L}^{m+1}_M\ and\ l^m \notin \mathcal{L}^m_M$$
\begin{equation} \label{eqn:eq1}
\Longrightarrow \mathcal{L}^m_M  \subset \mathcal{L}^{m+1}_M 
\end{equation}

\noindent
Since a link \textit{CLOSED} in \textit{ROUND} $m$, will remain \textit{CLOSED} for the subsequent \textit{ROUND}s, we have
$$if\ l \in \mathcal{L}^m_C,\ then\ l \in \mathcal{L}^k_C, \forall k \geq m+1$$
\begin{equation} \label{eqn:eq2}
\Longrightarrow \mathcal{L}^m_C  \subseteq \mathcal{L}^{m+1}_C 
\end{equation}

\noindent
Now from \eqref{eqn:eq1} and \eqref{eqn:eq2} we have
$$ \mathcal{L}^m_C \cup \mathcal{L}^m_M \subset \mathcal{L}^{m+1}_C \cup \mathcal{L}^{m+1}_M $$ 

\noindent
At all times, a link $l$ can be in one of the four states, i.e
$$ \forall m,\mathcal{L}^m_C \cup \mathcal{L}^m_M \cup \mathcal{L}^m_O \cup \mathcal{L}^m_H = \mathcal{L}$$

\noindent
From the above two argument,
$$ \mathcal{L}^{m+1}_O \cup \mathcal{L}^{m+1}_H \subset \mathcal{L}^{m}_O \cup \mathcal{L}^{m}_H $$

\noindent
Since the number of links in set $\mathcal{L}$ is finite, there exists a $t< \infty$, such that
$$ \mathcal{L}^{t}_O \cup \mathcal{L}^{t}_H = \lbrace \phi \rbrace $$

\noindent
Thus the algorithm terminates in finite number of \textit{ROUND}s and thus in finite time.
\end{IEEEproof}

\noindent
Let us assume that no two link prices are equal i.e $\forall i,j \in \mathcal{L}, \lambda_i \neq \lambda_j$ \\

\begin{lem}
$\forall i \in \mathcal{L}^m_C, \exists j \in \mathcal{L}^m_M : d(i,j) < K$.
\end{lem}

\begin{IEEEproof}
Assume that there is no such $j \in \mathcal{L}^m_M$ for some $i \in \mathcal{L}^m_C$.
Then the link $i$ would not have received any \textit{MARKED} link that interferes with it, in slot $S^{m-1}_M$ (Algorithm 2, At time $T^m_M$, lines 5, 6). Then this would imply that either link $i$ would be \textit{OPEN}ed or would be in \textit{CHECK}. i.e
$$ i \in \mathcal{L}^m_H \cup \mathcal{L}^m_O $$
$$ \Longrightarrow i \notin \mathcal{L}^m_C $$

\noindent
But this is a contradiction. Thus there exists a link $j \in \mathcal{L}^m_M$, such that $d(i,j) < K$.
\end{IEEEproof}

\begin{lem}
$\forall i \in \mathcal{L}^m_H, \exists j \in \mathcal{L}^m_O : \lambda_j > \lambda_i$
\end{lem}

\begin{IEEEproof}
Assume that there is no such $j \in \mathcal{L}^m_O$ for some $i \in \mathcal{L}^m_H$. Then this would imply that $\lambda_i > \lambda_j, \forall j \in \mathcal{L}^m_O$. Now since link $i \in \mathcal{L}^m_H$, we can say that a link $k$, such that $\lambda_k > \lambda_i$ and $d(k,i)=0$ was \textit{OPEN}ed at time $T^{m-1}_M$ (Algorithm 2, At time $T^m_M$, line 8). Since link $k$ was \textit{OPEN}ed at time $T^{m-1}_M$, $k \in \mathcal{L}^m_O$. Let
$$\lambda_i > \lambda_j, \forall j \in \mathcal{L}^m_O$$

\noindent
But we have $\lambda_k > \lambda_i$, Thus
$$\lambda_k > \lambda_j, \forall j \in \mathcal{L}^m_O$$
$$ \Longrightarrow k \notin \mathcal{L}^m_O$$
But this is a contradiction the statement that $k \in \mathcal{L}^m_O$. Thus $\forall i \in \mathcal{L}^m_H,\lambda_i < \lambda_j, \textrm{for some } j \in \mathcal{L}^m_O$
\end{IEEEproof}

\begin{lem}
At the beginning of a \textit{ROUND}, consider the globally highest priced link among links that are neither \textit{CLOSED} nor \textit{MARKED}. Such a link will not be in \textit{CHECK} state.
\end{lem}

\begin{IEEEproof}
Let
$$ \lambda^m_{max} = \max_{l \in \mathcal{L}, l \notin \mathcal{L}^m_M \cup \mathcal{L}^m_C}{\lambda_l}$$

\noindent
Equivalently,
$$ \lambda^m_{max} = \max_{l \in \mathcal{L}^m_O \cup \mathcal{L}^m_H}{\lambda_l}$$

\noindent
Or,
$$ \lambda^m_{max} = \max{(\max_{k \in \mathcal{L}^m_O}{\lambda_k},\max_{l \in \mathcal{L}^m_H}{\lambda_l})}$$

\noindent
Let
$$ i = arg \max_{l \in \mathcal{L}^m_H}{\lambda_l}$$

\noindent
It is evident that $i \in \mathcal{L}^m_H$. Thus from the previous claim, there exists a $j \in \mathcal{L}^m_O$ such that
$$ \lambda_j > \lambda_i$$

\noindent
i.e,
$$ \lambda_j > \max_{l \in \mathcal{L}^m_H}{\lambda_l}$$

\noindent
Also
$$ \max_{k \in \mathcal{L}^m_O}\lambda_k \geq \lambda_j$$
$$ \max_{k \in \mathcal{L}^m_O}\lambda_k > \max_{l \in \mathcal{L}^m_H}{\lambda_l}$$

\noindent
Hence
$$ \lambda^m_{max} = \max_{k \in \mathcal{L}^m_O}{\lambda_k}$$

\noindent
Thus $arg\ \lambda^m_{max} \in \mathcal{L}^m_O $.
\end{IEEEproof}

\begin{flushleft}
Let $\mathcal{L}_C$ be the set of links \textit{CHOSEN} by the centralized greedy algorithm. \\
Let the set $\mathcal{L}_C$ be ordered and indexed in the decreasing order of link price as $\lbrace l_1, l_2,...,l_v... \rbrace$. \\
Let the distributed greedy algorithm terminate after $t$ \textit{ROUND}s. Let $\mathcal{L}^{t+1}_M$ be the set of \textit{MARKED} links after the termination of the algorithm. \\
\end{flushleft}

\begin{lem}
The distributed greedy algorithm and the centralized greedy algorithm, schedule the same links.
\end{lem}

\begin{IEEEproof}
We need to prove that every link \textit{CHOSEN} by the centralized greedy algorithm is \textit{MARKED} by the distributed greedy algorithm, by the time it terminates i.e. $\mathcal{L}_C \subseteq  \mathcal{L}^{t+1}_M$. We will prove the above claim via induction. 

\noindent
\textit{Induction statement:} If links $l_1,l_2,...,l_k \in \mathcal{L}_C$ then $l_1,l_2,...,l_k \in \mathcal{L}^{t+1}_M$.

\noindent
\textit{Basis:} To show the statement holds for the globally maximum priced link.\\
Let link $l_1 \in \mathcal{L}_C$ be the globally maximum priced link. Thus this link will also be a local maximum among interfering links in a $(K + 1)$-hop neighbourhood. Thus this link will be \textit{MARKED} after the $1^{st}\ ROUND$. i.e 
$$l_1 \in \mathcal{L}^{2}_M$$

\noindent
A link which is \textit{MARKED}, will continue to remain so.
$$\therefore l_1 \in \mathcal{L}^{t+1}_M$$

\noindent
Let us define 
$$\mathcal{I}(y) = \lbrace l \in  \mathcal{L}:d(l, y) < K \rbrace$$

\noindent
as the set of links that interfere with link $y$. Let 
$$ \mathcal{L}^{k+1} = \mathcal{L} - \cup^k_{i=1}(l_i \cup \mathcal{I}(l_i))$$

\noindent
be the set of links left after links $\lbrace l_1,l_2,...l_k \rbrace$ are \textit{CHOSEN}. Let
\begin{equation} \label{eqn:eq3}
\forall l \in \mathcal{L}^{k+1}, \mathcal{P}(l) = \lbrace l^{'} \in \mathcal{L}^{k+1}:d(l,l^{'}) < K, \lambda_{l^{'}} > \lambda_{l} \rbrace
\end{equation}

\noindent
It is obvious that for link $L_{k+1}$ to be \textit{CHOSEN}, $\mathcal{P}(l_{k+1})= \lbrace \phi \rbrace$ .  \\

\par \noindent
\textit{Inductive step:} If  $l_1,l_2,...,l_k \in \mathcal{L}^{t+1}_M$ given that  $l_1,l_2,...,l_k \in \mathcal{L}_C$, then if $l_{k+1} \in \mathcal{L}_C$ then $l_{k+1} \in \mathcal{L}^{t+1}_M$. 

\noindent
Since $l_1,l_2,...,l_k \in \mathcal{L}^{t+1}_M$, for each $i \in \lbrace 1,2,...,k \rbrace$
$$ \exists m_i \leq t+1 : l_i \in \mathcal{L}^{m_i}_M, l_i \notin \mathcal{L}^{s}_{M} \textrm{ \textit{for} } s < m_i$$

\noindent
Since \textit{MARKED} links always remains in the same state,
$$l_i \in \mathcal{L}^{r}_M, \quad  \forall m_i \leq r \leq t+1 \quad  $$

\noindent
Let
$$\forall l \in \mathcal{L}^{m}_O, \mathcal{P}^{m}(l) = \lbrace l^{'} \in \mathcal{L}^m_O : d(l,l^{'}) < K, \lambda_{l^{'}} > \lambda_{l} \rbrace$$

\noindent
We note that link $l$ is \textit{MARKED} in \textit{ROUND} $m$, if $\mathcal{P}^{m}(l)= \lbrace \phi \rbrace $. Let
$$ m^{'} = \max^{k}_{i=1}{m_i}$$

\noindent
It is easy to see that before \textit{ROUND} $m^{'}$, links $\lbrace l_1,l_2,...l_k \rbrace$ are \textit{MARKED} and correspondingly, all the links that interfere with these links are \textit{CLOSED}. Thus
$$ \mathcal{L}^{m^{'}}_O \subseteq \mathcal{L} - \cup^k_{i=1}(l_i \cup \mathcal{I}(l_i))$$
\begin{equation} \label{eqn:eq4}
\mathcal{L}^{m^{'}}_O \subseteq \mathcal{L}^{k+1} 
\end{equation}

\noindent
Now,
$$ \mathcal{P}^{m^{'}}(l_{k+1}) = \lbrace l^{'} \in \mathcal{L}^{m^{'}}_O : d(l_{k+1},l^{'}) < K, \lambda_{l^{'}} > \lambda_{l_{k+1}} \rbrace$$
From \eqref{eqn:eq3} and \eqref{eqn:eq4}, we can say that
$$\mathcal{P}^{m^{'}}(l_{k+1}) \subseteq \mathcal{P}(l_{k+1}) $$

\noindent
Since $l_{k+1} \in \mathcal{L}_C,\ \mathcal{P}(l_{k+1})= \lbrace \phi \rbrace $. Thus
$$ \mathcal{P}^{m^{'}}(l_{k+1}) \subseteq \lbrace \phi \rbrace $$
$$ \Rightarrow  \mathcal{P}^{m^{'}}(l_{k+1}) = \lbrace \phi \rbrace  $$

\noindent
Now, since the algorithm terminates after $t$ \textit{ROUND}s, 
$$\mathcal{L}^{t+1}_O \cup \mathcal{L}^{t+1}_H = \lbrace \phi \rbrace $$
$$\therefore l_{k+1} \notin \mathcal{L}^{t+1}_O$$
$$ \Rightarrow m^{'} \neq t+1$$

\noindent
Thus link $l_{k+1}$ gets \textit{MARKED} in \textit{ROUND} $m^{'} \leq t$. Thus link $l_{k+1}$ gets \textit{MARKED} before the algorithm terminates. Therefore, 
$$\forall l \in \mathcal{L}_C,\ l \in \mathcal{L}^{t+1}_M$$
$$ \Rightarrow  \mathcal{L}_C \subseteq \mathcal{L}^{t+1}_M$$

\noindent
Now, let us assume that LHS is a strict subset of RHS, i.e
$$ \mathcal{L}_C \subset \mathcal{L}^{t+1}_M $$

\noindent
Then there exists a link $l_i$ such that $l_i \notin \mathcal{L}_C$ but $l_i \in \mathcal{L}^{t+1}_M$. Since $l_i \in \mathcal{L}^{t+1}_M$, we can say that
$$d(l,l_i) \geq K, \forall l \in \mathcal{L}^{t+1}_M$$

\noindent
Since $ \mathcal{L}_C \subset \mathcal{L}^{t+1}_M $, 
$$d(l,l_i) \geq K, \forall l \in \mathcal{L}_C$$

\noindent
But then if the above was true, then $l_i \in \mathcal{L}_C$. But this is a contradiction. Thus LHS can not be a strict subset of RHS.
$$ \Rightarrow  \mathcal{L}_C = \mathcal{L}^{t+1}_M$$
\end{IEEEproof}

Here, we would like to derive closed form expression for the worst case run time of the distributed greedy heuristic. Let us assume $T$ to be the worst case time for a \textit{ROUND} to complete. Let $|\mathcal{L}|$ be the number of links in the wireless network. \\

\newpage

\begin{lem}
The worst case termination time for the algorithm under the $K$-hop link interference model is $\frac{|\mathcal{L}| \cdot T}{K}$.
\end{lem}

\begin{IEEEproof}
If we assume $K$-hop link interference model, on a wireless network with $|\mathcal{L}|$ number of links, it is easy to see that the centralized greedy algorithm can schedule at-most $\frac{|\mathcal{L}|}{K}$ links. Using the result from previous lemma, we can argue that the distributed greedy algorithm too can schedule a maximum of $\frac{|\mathcal{L}|}{K}$ links. The algorithm sets at-least one link as \textit{MARKED} in each \textit{ROUND}. Since the maximum number of links schedule will be $\frac{\mathcal{L}|}{K}$, we can say that the algorithm terminates after at-most $\frac{|\mathcal{L}|}{K}$ \textit{ROUND}s. Since each \textit{ROUND} takes a worst case time $T$, the worst case termination time for the algorithm is $\frac{|\mathcal{L}| \cdot T}{K}$.
\end{IEEEproof}

\section{\large{Conclusion}}
\label{conclude}
The scheduling problem is known to be a bottleneck in the cross-layer optimization approach. The interference constraints were modeled using the $K$-hop link interference model. Under the assumption that each node transmits at a fixed power level (which can be different for different nodes), the optimal scheduling problem is a weighted matching problems with constraints determined by the $K$-hop interference model. We explored the greedy heuristic because it is amenable to distributed implementation. In this paper, we have come up with a distributed greedy heuristic and have shown that it performs exactly as the centralized greedy heuristic. In future work, we plan to find how close the distributed greedy heuristic is to the optimal solution. We would also like to explore the scope of distributed greedy heuristics for other network problems. This distributed greedy algorithm was implemented as a part of full-fledged distributed protocol for aggregate utility maximization. During the incorporation of the above mentioned algorithm into the protocol, several new issues were to be tackled, like implementation of a reliable message passing mechanism. Many issues like this require careful analysis and quantification. We do not discuss these problems in this paper due to lack of space.

\IEEEtriggeratref{1}

\bibliographystyle{IEEEtran}
\bibliography{distributed_greedy_final}

\begin{thebibliography}{10}
\providecommand{\url}[1]{#1}
\csname url@samestyle\endcsname
\providecommand{\newblock}{\relax}
\providecommand{\bibinfo}[2]{#2}
\providecommand{\BIBentrySTDinterwordspacing}{\spaceskip=0pt\relax}
\providecommand{\BIBentryALTinterwordstretchfactor}{4}
\providecommand{\BIBentryALTinterwordspacing}{\spaceskip=\fontdimen2\font plus
\BIBentryALTinterwordstretchfactor\fontdimen3\font minus
  \fontdimen4\font\relax}
\providecommand{\BIBforeignlanguage}[2]{{%
\expandafter\ifx\csname l@#1\endcsname\relax
\typeout{** WARNING: IEEEtran.bst: No hyphenation pattern has been}%
\typeout{** loaded for the language `#1'. Using the pattern for}%
\typeout{** the default language instead.}%
\else
\language=\csname l@#1\endcsname
\fi
#2}}
\providecommand{\BIBdecl}{\relax}
\BIBdecl

\bibitem{tassiulas}
{L. Tassiulas} and {A. Ephremides}, ``Stability properties of constrained
  queueing systems and scheduling policies for maximum throughput in multihop
  radio networks,'' in \emph{IEEE Trans. on Automatic Control,
  37(12):1936–1948}, December 1992.

\bibitem{brar}
{G. Brar}, {D.M Blough}, and {P. Santi}, ``The scream approach for efficient
  distributed scheduling with physical interference in wireless mesh
  networks,'' in \emph{Distributed Computing Systems, ICDCS}, 2008.

\bibitem{gupta}
{P. Gupta} and {P. R. Kumar}, ``The capacity of wireless networks,'' in
  \emph{IEEE Transactions on Information Theory}, December 2000.

\bibitem{sharma}
{G. Sharma}, {R.R. Mazumdar}, and {N.B. Shroff}, ``On the complexity of
  scheduling in wireless networks,'' in \emph{Proc. of ACM MobiCom}, September
  2006.

\bibitem{lin}
{X. Lin} and {N. B. Shroff}, ``The impact of imperfect scheduling on
  cross-layer rate control in multihop wireless networks,'' in
  \emph{Proceedings of IEEE INFOCOM}, Miami, FL, March 2005.

\bibitem{chaporkar}
{P. Chaporkar}, {K. Kar}, and {S. Sarkar}, ``Throughput guarantees in maximal
  scheduling in wireless networks,'' in \emph{Proceedings of 43d Annual
  Allerton Conference on Communication, Control and Computing}, Monticello,
  September 2005.

\bibitem{wu}
{X. Wu} and {R. Srikant}, ``Regulated maximal matching: A distributed
  scheduling algorithm for multi-hop wireless networks with node exclusive
  spectrum sharing,'' in \emph{Proceedings of the IEEE Conference on Decision
  and Control}, Seville, Spain, December 2005.

\bibitem{joo}
{C. Joo} and {N. Shroff}, ``Performance of random access scheduling schemes in
  multi-hop wireless networks,'' in \emph{IEEE INFOCOM}, Anchorage, AK, May
  2007.

\bibitem{lin1}
{X. Lin} and {S. Rasool}, ``Constant-time distributed scheduling policies for
  ad hoc wireless networks,'' in \emph{Proceedings of the IEEE Conference on
  Decision and Control}, San Diego, CA, December 2006.

\bibitem{wu1}
{X. Wu}, {R. Srikant}, and {J. R. Perkins}, ``Queue-length stability of maximal
  greedy schedules in wireless networks,'' in \emph{Proc. Workshop on
  Information Theory and Applications}, UCDS, February 2006.

\bibitem{neely}
{M. J. Neely}, {E. Modiano}, and {C. E. Rohrs}, ``Dynamic power allocation and
  routing for time varying wireless networks,'' in \emph{IEEE INFOCOM}, San
  Francisco, April 2003.

\bibitem{chiang}
{M. Chiang}, ``To layer or not to layer: Balancing transport and physical
  layers in wireless multihop networks,'' in \emph{IEEE INFOCOM}, Hong Kong,
  March 2004.

\bibitem{soldati}
{P. Soldati}, {B. Johansson}, and {M. Johansson}, ``Proportionally fair
  allocation of end-to-end bandwidth in stdma wireless networks,'' in
  \emph{Proceedings of the 7th ACM international symposium on Mobile ad hoc
  networking and computing}, New York, 2006.

\bibitem{gro}
{J. Groonkvist}, ``Assignment methods for spatial reuse tdma,'' in
  \emph{Proceedings of the 1st ACM international symposium on Mobile ad hoc
  networking and computing}, New Jersey, 2000.

\bibitem{kim}
{J. Kim}, {X. Lin}, and {N. B. Shroff}, ``Locally-optimized scheduling and
  power control algorithms for multi-hop networks under sinr interference
  models,'' in \emph{Proceedings of the 5th International Symposium on Modeling
  and Optimization in Mobile, Ad Hoc, and Wireless Networks (WiOpt 2007)},
  Limassol, Cyprus, April 2007.

\bibitem{jain}
{K. Jain}, {J. Padhye}, {V.N. Padmanabhan}, and {L. Qiu}, ``Impact of
  interference on multi-hop wireless network performance,'' in \emph{Proc. of
  IEEE MobiCom}, September 2003.

\end{thebibliography}

\end{document}